\documentclass[nofootinbib,showpacs,preprintnumbers,amsmath,amssymb,floatfix]
{revtex4-1}
\usepackage{graphicx}
\usepackage{epsf}
\usepackage{wrapfig}

\begin{document}


\title{Unpolarized and spin-dependent DIS structure functions in Double-Logarithmic Approximation}

\vspace*{0.3 cm}

\author{B.I.~Ermolaev}
\affiliation{Ioffe Physico-Technical Institute, 194021
 St.Petersburg, Russia}
\author{S.I.~Troyan}
\affiliation{St.Petersburg Institute of Nuclear Physics, 188300
Gatchina, Russia}

\begin{abstract}
We demonstrate how to calculate perturbative components of the structure functions $F_1$ (for unpolarized DIS) and $g_1$ (spin-dependent DIS) in
Double-Logarithmic Approximation, studying separately the cases of fixed and running QCD coupling. We show that as long as
only ladder graphs are accounted for (throughout the talk we use the Feynman gauge for virtual gluons) there is no difference at all
between $F_1$ and $g_1$. However, accounting for contributions of non-ladder graphs brings an essential difference between them. Applying the Saddle-Point
method to the obtained expressions for $F_1$ and $g_1$ allows us to arrive at their small-$x$ asymptotics. The both asymptotics are of the Regge kind but with
different intercepts.  The intercept of $F_1$ proved to be greater than unity, so it is a new contribution to Pomeron.
Finally, we discuss the applicability region of the Regge asymptotics and demonstrate that inappropriate
replacement of $F_1$ and $g_1$ by their
asymptotics outside the applicability region can lead to introducing phenomenological Pomerons for both unploarized and spin-dependent
processes.
\end{abstract}


\maketitle

\section{Introduction}

According to QCD factorization, structure functions describing Deeply Inelastic Scattering can be regarded as convolutions of
perturbative and non-perturbative contributions. In the present talk we consider the perturbative components only. In particular,
we consider perturbative components of the DIS structure functions $F_1$ and $g_1$.
They can be calculated in fixed orders in the QCD coupling or, alternatively, with total resummation of contributions to
all orders in in $\alpha_s$. The latter can be done only approximately, with selecting most important contributions
in every order  in $\alpha_s$ and constructing evolution equations accounting for such contributions.
The most important contributions  are different in different kinematics. Selection of them is expressed in terms of
orderings of momenta of virtual partons.  They look very simple when the standard Sudakov parametrization\cite{sud} is used
for momenta $k_i (i= 1,2,...)$ of virtual partons:

\begin{equation}\label{sud}
k_i = \alpha_i q^{\prime} + \beta_i p{\prime} + k_{i \perp},
\end{equation}
where $q^{\prime}$ and $p^{\prime}$ are light-cone momenta, each made of the photon momentum $q$ and the initial parton momentum $p$
while $k_{i \perp}$ are the components transversal to the plane formed by $p$ and $q$. The invariant energy $w = 2pq$ is presumed
to be the largest invariant involved.
First of all, there is the DGLAP ordering:

\begin{equation}\label{dglaporder}
\beta_1 \sim \beta_2 \sim... \sim 1,~~~ \mu^2 \ll k^2_{1 \perp} \ll  k^2_{2 \perp} \ll... \ll Q^2,
\end{equation}
where $-Q^2 = q^2$ and $\mu^2$ is a mass scale. For instance, it can be
the factorization scale. We keep the standard DGLAP numeration of partonic ladder rungs from the bottom to the top in involved Feynman graphs.
This ordering means that the DGLAP equations\cite{dglap} sum logarithms of $Q^2$ to all orders in $\alpha_s$ and do not account for logs of $x$.
As a result, DGLAP is designed for work in kinematics $x \sim 1$ and $Q^2 \gg \mu^2$ for reactions with both
unpolarized and polarized partons, so it describes both $F_1$ and $g_1$ at large $x$.  In contrast, the
BFKL ordering is

\begin{equation}\label{bfklorder}
1 \gg \beta_1 \gg \beta_2 \gg...,~~~ \mu^2 \sim k^2_{1 \perp} \sim  k^2_{2 \perp} \sim....
\end{equation}

So, the BFKL equation\cite{bfkl} accounts for logs of $x$.
and does dot deal with logarithms of $Q^2$. As a result, BFKL is tailored for work in kinematic region of small $x$ and small $Q^2$.
It sums leading logarithms and contributes to unpolarized processes only, i.e. BFKL contributes to description of $F_1$ but not
$g_1$. The perturbative series for $F_1$ in Leading Logarithmic Approximation (LLA) looks as follows:

\begin{equation}\label{bfklser}
F_1^{LL} = \delta (x-1) + (1/x) \left[1 + c_1 \alpha_s \ln (1/x) + c_2 (\alpha_s \ln (1/x))^2 + ...\right],
\end{equation}
where $c_r$ are numerical factors.
Alternatively, both logs of $Q^2$ and logs of $x$ are accounted for when
the Double-Logarithmic Ordering\cite{ggfl} is used:

\begin{equation}\label{dlorder}
1 \gg \beta_1 \gg \beta_2 \gg...,~~~ \mu^2 \sim k^2_{1 \perp}/\beta_1 \ll  k^2_{2 \perp}/\beta_2 \ll....
\end{equation}

This ordering makes possible to account for logs of $x$ and $Q^2$ in Double-Logarithmic Approximation (DLA) for both
unpolarized and spin-dependent processes and therefore both $F_1$ and $g_1$ can be calculated in DLA. The DL
perturbative series for both $g_1$ and $F_1$ looks as follows:

\begin{eqnarray}\label{dlser}
F_1^{DL} &=& \delta (x-1) + c'_1 \alpha_s \ln^2 (1/x) + c'_2 (\alpha_s \ln^2 (1/x))^2 + ...,
\\ \nonumber
g_1^{DL} &=& \delta (x-1) + \tilde{c}_1 \alpha_s \ln^2 (1/x) + \tilde{c}_2 (\alpha_s \ln^2 (1/x))^2 + ...
\end{eqnarray}
where $c'_r, \tilde{c}_r$ are numerical factors. The overall factor $1/x$ in Eq.~(\ref{bfklser}) is huge at small $x$, so the
DL contribution $F_1^{DL}$ of Eq.~(\ref{dlser}) looks negligibly small compared to $F_1^{LL}$. In the present talk we demonstrate
that this impression is false.

\section{Calculating the structure functions $F_1$ and $g_1$ in DLA}

In order to calculate $F_1$ and $g_1$ in DLA $F_1$ and $g_1$ in DLA we construct and solve Infra-red Evolution Equations (IREEs).
This method was suggested by L.N.~Lipatov. The basic idea is to introduce an infra-red cut-off $\mu$ in the transverse
momentum space and trace evolution with respect to $\mu$. The key-stone idea here is factorization of DL
contributions of partons with minimal transverse momenta, which was proved by
V.N.~Gribov in the QED context. History and details of application of the method
to DIS can be found in Ref.~\cite{egtg1sum}. It is convenient to begin with calculating amplitudes of elastic
Compton scattering off a quark and a gluon, which we denote $A_q$ and $A_g$ respectively, and obtain
$F_1^{(q,g)}$ and $g_1^{(q,g)}$ from them with Optical theorem:

\begin{eqnarray}\label{opt}
F_1^q   &=&  \frac{1}{2\pi} \Im A_q^{(+)}, ~~~~~~F_1^g   =  \frac{1}{2\pi} \Im A_g^{(+)}, \\
\nonumber
g_1^q   &=&  \frac{1}{2\pi} \Im A_q^{(-)}, ~~~~~~g_1^g   =  \frac{1}{2\pi} \Im A_g^{(-)},
\end{eqnarray}
where the signature amplitudes $A_q^{(\pm)}$ and $A_g^{(\pm)}$ defined as follows:

\begin{equation}\label{pm}
A_q^{(\pm)}(w,Q^2) = A_q (w,Q^2) \pm A_q (-w,Q^2),~~~A_g^{(\pm)}(w,Q^2) = A_g (w,Q^2) \pm A_g (-w,Q^2).
\end{equation}

It is convenient to express $A_q$ and $A_g$ through the Mellin transform:

 \begin{equation}\label{mellin}
A^{(\pm)}_{q,g}(w/\mu^2, Q^2/\mu^2) = \int_{- \imath \infty}^{\imath \infty}
\frac{d \omega}{2 \pi \imath} \left(w/\mu^2\right)^{\omega}\xi^{(\pm)}(\omega) F^{(\pm)}_{q,g}(\omega, Q^2/\mu^2)
= \int_{- \imath \infty}^{\imath \infty}
\frac{d \omega}{2 \pi \imath} e^{\omega \rho}\xi^{(\pm)}(\omega) F^{(\pm)}_{q,g}(\omega, y),
\end{equation}
where we have introduced the signature factor $\xi^{(\pm)}(\omega) = -\left(e^{- \imath \omega} \pm 1\right)/2$ and
the logarithmic variables $\rho, y$ (using the standard notation $w = 2pq$):
\begin{equation}\label{y12}
\rho = \ln (w/\mu^2),~~y = \ln (Q^2/\mu^2).
 \end{equation}
 In what follows we will address $F_q^{(\pm)}, F_g^{(\pm)}$ as Mellin amplitudes and will use the same form of the Mellin transform for other amplitudes as well. Constructing IREEs for
the Compton amplitudes $F_q^{(\pm)}$ and $F_q^{(\pm)}$ is
identical, so we keep generic notations $F_q$ and $F_g$ for them without the signature superscripts.
Technology of composing and solving IREEs in detail can be found in Ref.~\cite{egtg1sum,etf1}.  As the first step, we
construct IREEs involving
$F_{q,g}$ and auxiliary amplitudes $h_{rr'}$. They are related to the parton-parton amplitudes $f_{rr'}$:
$h_{rr'} = \frac{1}{8 \pi^2} f_{rr'}$,
with $r,r' = q,g$. So, we obtain the following IREEs:

 \begin{eqnarray}\label{eqf}
\left[\partial/\partial y + \omega \right] F_{q}(\omega,y) &=&
F_{q} (\omega, y) h_{qq} (\omega) + F_{g} (\omega, y) h_{gq} (\omega),
\\ \nonumber
\left[\partial/\partial y + \omega \right] F_{g}(\omega, y) &=&
F_{q} (\omega, y) h_{qg} (\omega) + F_{g} (\omega, y) h_{gg} (\omega),
\end{eqnarray}

Solving Eqs.~(\ref{eqf}), we express $F_{q,g}$ in terms of auxiliary amplitudes $h_{rr'}$ which
can be found by the same method. Explicit expressions for them can be found
in \cite{egtg1sum,etf1}. Substituting them in solutions of Eqs.~(\ref{eqf})
allows us to arrive at explicit expressions for $F_{q,g}$ and then obtain $F_1$ and $g_1$ in DLA.

\section{Small-$x$ asymptotics of $F_1$ and $g_1$}

Pushing $x \to 0$ and applying Saddle-Point method to the expressions for $F_1$ and $g_1$, we arrive at
their small-$x$ asymptotics. They both are of the Regge kind, though with different stationary points $\omega^{(\pm)}_0$:

\begin{eqnarray}\label{asympt}
g_1 \sim \frac{\Pi\left(\omega^{(-)}_0\right)}{\ln^{3/2}(1/x)} x^{-\omega^{(-)}_0} \left(\frac{Q^2}{\mu^2}\right)^{\omega^{(-)}_0/2},
\\ \nonumber
F_1 \sim \frac{\Pi\left(\omega^{(+)}_0\right)}{\ln^{3/2}(1/x)} x^{-\omega^{(+)}_0} \left(\frac{Q^2}{\mu^2}\right)^{\omega^{(+)}_0/2},
\end{eqnarray}
where we again introduced the signature notations $(\pm)$.
Explicit expressions of the factors $\Pi(\omega^{(\pm)}_0)$ depend on the type of
QCD factorization (see Refs.~\cite{egtg1sum,etf1} for detail). In Regge theory, $\omega^{(\pm)}_0$ are called intercepts. They control
the $x$-dependence of the structure functions. Intercept $\omega^{(-)}_0$ of $g_1$ was calculated in Ref.~\cite{egtg1} and
$\omega^{(+)}_0$ was obtained in Ref.~\cite{etf1}. When the running $\alpha_s$ effects are accounted for, the intercepts are:

\begin{equation}\label{omegazero}
\omega^{(-)}_0= 0.86,~~~~~~\omega^{(+)}_0 = 1.07.
\end{equation}

It is interesting to notice that the intercept $\omega^{(-)}_0$ is in good agreement with the result $\omega^{(-)}_0 = 0.88 \pm 0.17$
obtained in Ref.~\cite{koch} by extrapolating the HERA data to the region of $x \to 0$. Intercept $\omega^{(+)}_0 > 1$, so this Reggeon
is a new contribution to Pomeron. Throughout the talk we will address it as DL Pomeron.
Despite its value is pretty close to the NLO BFKL Pomeron intercept, DL and BFKL Pomerons have nothing in common: BFKL equation sums
leading logarithms whereas IREEs of Eq.~(\ref{eqf}) deal with double logarithms.

Another interesting observation is that the intercepts $\omega^{(+)}_0$ and $\omega^{(-)}_0$ coincide as long as DL contributions of non-ladder
Feynman graphs\footnote{the terms "ladder/non-ladder contributions" are gauge-dependent. We use them in regard of the
Feynman gauge.} are neglected. In this case $\omega^{(+)}_0 = \omega^{(-)}_0 = 1.25$. Non-ladder graphs contributions diminish them both
but their impact on $\omega^{(-)}_0$  is greater than on $\omega^{(+)}_0$. Effect of difference in such
impacts was first noticed in Ref.~\cite{nest} in the
QED context.

Eq.~(\ref{asympt}) manifests that Regge asymptotics of  are represented by simple and elegant expressions in contrast
to the parent amplitudes/structure functions. However, these asymptotics should be used within their applicability regions.  Keeping a general notation $F$ for $F_1$ and $g_1$ and
denoting $\widetilde{F}$ their asymptotics, we introduce their ratio $R$ as follows:

\begin{equation}\label{r}
 R (x, Q^2) = \widetilde{F}(x,Q^2)/F (x,Q^2).
\end{equation}

Obviously, the asymptotics reliably represent their parent structure functions when $R \approx 1$. Let us fix $Q^2 = 10~$GeV$^2$ and study
the $x$-dependence of $R$. Numerical calculations yield that $R >0.9$ at $x < x_{max}$, with

\begin{equation}\label{xmas}
x_{max} = 10^{-6}.
\end{equation}

Nevertheless, it is well-known that in practice the Regge asymptotics have been used at $x \gg x_{max}$. Doing so leads to artificial increase of the intercepts.
Indeed, let us assume that the model Pomeron $x^{-a}$ is used at $x = x_1= 10^{-4}$.
It is supposed to represent $F$ and therefore $x_1^{-a} \approx F$. On the other hand, Eq.~(\ref{xmas}) states that $\left(x_{max}\right)^{-\omega^{(+)}_0}$. Equating these expressions, we arrive at

\begin{equation}\label{xaomega}
x_1^{-a} \approx \left(x_{max}\right)^{-\omega^{(+)}_0},
\end{equation}
which leads to

\begin{equation}\label{aplus}
a = \frac{3}{2}~ \omega^{(+)}_0 \approx 1.6,
\end{equation}
which means that the model Pomeron is hard.
Applying the above reasoning at the same value $x_1 = 10^{-4}$ to the spin-dependent Reggeon in Eq.~(\ref{asympt}), with the intercept $\omega^{(-)}_0 = 0.86$, makes easy to arrive at
a new "Reggeon" with the ”intercept”

\begin{equation}\label{aminus}
a^{(-)} = \frac{3}{2}~\omega_0^{(-)} \approx 1.3
\end{equation}
and obtain thereby a fictitious spin-dependent Pomeron.
These examples manifest that using the small-$x$ asymptotics outside their applicability region can certainly lead to introducing
hard Pomerons for both  unpolarized and spin-dependent DIS, though without theoretical grounds.

\section{Conclusion}

In the present talk we have demonstrated how to calculate the structure functions $F_1$ and $g_1$ in DLA and how to
calculate their small-$x$ asymptotics. It turned out that the both asymptotics are of the Regge form but their intercepts
are different. They coincide when only the ladder Feynman graphs are accounted for but impact of double logarithms from
non-ladder graphs brings different  contributions to these intercepts. As a result, the intercept $\omega^{(-)}_0$
of the $g_1$ asymptotics in Eq.~(\ref{asympt}) is less than unity while the intercept $\omega^{(+)}_0$ of $F_1$-asymptotics
is a bit greater than unity
and therefore this Reggeon is a new Pomeron. Obviously, it has nothing in common with the BFKL Pomeron.

We also fixed in Eq.~(\ref{xmas}) the maximal value of $x$ where the small-$x$ asymptotics of $g_1$ and $F_1$ can be used instead of
the parent structure functions. Exploiting this estimate, we conclude
that widespread substitution of scattering amplitudes or structure functions by their Regge asymptotics
at experimentally available energies leads to introducing phenomenological Pomerons for both unpolarized and spin-dependent processes.
The larger $x$ for using the Regge asymptotics are chosen, the greater phenomenological intercepts are required.

\section{Acknowledgement}

B.I.~Ermolaev is grateful to Organizing Committee of the workshop DSPIN-19 for financial support.

\end{document}